\documentclass[aps,pra,showpacs,noshowkeys]{revtex4}
\usepackage{epsfig}
\usepackage{graphicx}
\usepackage{bm}
\usepackage{amssymb,amsmath,euscript,esint,mathtools}
\begin{document}
\newcommand{\Br}[1]{(\ref{#1})}
\newcommand{\Eq}[1]{Eq.\ (\ref{#1})}
\newcommand{\frc}[2]{\raisebox{1pt}{$#1$}/\raisebox{-1pt}{$#2$}}
\newcommand{\frcc}[2]{\raisebox{0.3pt}{$#1$}/\raisebox{-0.3pt}{$#2$}}
\newcommand{\frccc}[2]{\raisebox{1pt}{$#1$}\big/\raisebox{-1pt}{$#2$}}
\title{$\bm{\mathcal{P}},\bm{\mathcal{T}}$-odd Faraday effect as a tool for observation of CP violation in Standard Model}
\author{D. V. Chubukov$^{1}$ and L. N. Labzowsky$^{1,2}$}
\affiliation{$^1$St. Petersburg State University, Ulyanovskaya 3,  198504, St. Petersburg, Petrodvorets, Russia \\
$^2$Petersburg Nuclear Physics Institute, 188300, St. Petersburg, Gatchina, Russia
}

\begin{abstract}
It is proposed to employ the $\mathcal{P},\mathcal{T}$-odd Faraday effect, i.e. rotation of the polarization plane of the light propagating through a medium in presence of the electric field, as a tool for observation of $\mathcal{P},\mathcal{T}$-odd effects caused by CP violation within the Standard Model. For this purpose the vapors of heavy atoms like Tl, Pb, Bi are most suitable. Estimates within the Standard Model show: provided that applied field is about $10^5 \;V/cm$ and the optical length can be as large as 70000 km, the rotation angle may reach the value corresponding to the recently observable values ($10^{-9} \; rad$). These estimates demonstrate that the $\mathcal{P},\mathcal{T}$-odd Faraday effect observations may effectively compete with the recent measurements of the electron spin rotation in an external electric field, performed with diatomic molecules. These measurements exclude the $\mathcal{P},\mathcal{T}$-odd effects at the level 9 orders of magnitude higher than the predictions of the Standard Model.
\end{abstract}
\maketitle
The CP-violation effects within the Standard Model (SM) manifest (due to CPT-conservation) the existence of the $\mathcal{T}$-noninvariant interactions in the nature. However up to now CP violation was observed only in exotic reactions with K-mesons \cite{Chris64}. The existence of the Electric Dipole Moment (EDM) for any (not real neutral) particle or for any closed system containing such particles (atoms, molecules, nuclei) also would mean the existence of $\mathcal{P},\mathcal{T}$ (i.e. $\mathcal{P}$- and $\mathcal{T}$)-odd interactions. A search for EDMs started in \cite{Pur50} for neutrons, in \cite{San65} for atoms and became the most long-standing problem in fundamental physics. It was found theoretically that the electron EDM in heavy atoms can be strongly enhanced compared to the EDM of free electrons \cite{San65}, \cite{Flam76}. Even stronger enhancement for the nuclear EDM in heavy diatomic molecules was predicted in \cite{San67} where the molecules with closed electron shells were considered. In \cite{Lab78} it was found that in heavy diatomic molecules with open electron shells very strong enhancement (up to 1 billion times) can arise for $\mathcal{P}$-odd effects due to the existence of the $\Lambda$-doubling effect, absent for the closed shell molecules. In \cite{Sush78} the same enhancement in diatomic molecules was predicted for the electron EDM and in \cite{Gor79} a $\mathcal{P},\mathcal{T}$-odd interaction between the electron and the nucleus in diatomic molecules was first considered. In \cite{Gor79} it was also demonstrated that the $\mathcal{P},\mathcal{T}$-odd electron-nucleus interaction effect cannot be distinguished from the electron EDM effect in any experiment with any particular atom or molecule. According to \cite{Bon15} this can be done only in the series of experiments with the Highly Charged Ions (HCI) due to the different dependence of the two effects on the nuclear charge $Z$. The most restrictive bounds for the electron EDM were established in experiments with Tl atom ($d_e<1.6\cdot 10^{-27} \, e\, cm$) \cite{Reg02}, YbF molecule ($d_e<10.5\cdot 10^{-28} \, e\, cm$) \cite{Hud11} and ThO molecule ($d_e< 8.7 \cdot 10^{-29}\, e\, cm$) \cite{ACME13}. For extraction of $d_e$ values from the experimental data the theoretical calculations of the enhancement coefficient are necessary. These calculations become especially sophisticated for molecules. In case of ThO these calculations were given by two groups \cite{Skrip13}, \cite{Fleig14}; results in \cite{Skrip13}, \cite{Fleig14} differ from each other by $10 \%$.

What concerns the theoretical predictions for the value $d_e$ within the Standard Model, the situation is rather uncertain. It was early understood that the CP-violating effects within the Standard Model can arise only via the phase factor in Cabibbo-Kobayashi-Maskawa (CKM) matrix. In particular, electron EDM can arise if 3-loop vertex is introduced including one quark loop which provides CKM phase \cite{Hoog90}. In \cite{Hoog90} the loop integrals were calculated numerically with the Glashow-Iliopoulos-Maiani (GIM) mechanism taken into account. The value obtained in \cite{Hoog90} for the electron EDM was $d_e\approx 10^{-38} \, e\, cm$. However later in \cite{Pos91} it was proven that the total result at 3-loop level should be exactly zero due to the cancelations between different quark flavors. In \cite{Pos91} it was suggested that with additional 4th loop containing one gluon exchange the result will become nonzero. In \cite{Pos14} it was claimed that the largest ``benchmark'' contribution to the  $\mathcal{P},\mathcal{T}$-odd effects in atomic systems comes from the two-photon exchange between electron and nucleus, containing one CP-violating vertex. This would give the result $d_e^{eqv,2\gamma}\approx 10^{-38} \,e\,cm$ for the ``equivalent'' EDM, i.e. the EDM value that would lead to the same linear Stark shift. In the same paper the ``ordinary'' electron EDM effect was estimated as $d_e\approx 10^{-44} \,e\,cm$. This estimate was given exclusively on the basis of GIM, without numerical calculations. The contribution of the gluon exchange was expressed via the factor $\frc{\alpha_s}{4\pi}\approx \frc{1}{10}$, where $\alpha_s$ is the strong interaction constant. Multiplying this factor by the $ 10^{-38} \, e\, cm$ \cite{Hoog90}, we would obtain $d_e\approx 10^{-39} \,e\,cm$. The gap between the values $d_e\approx 10^{-39} \, e\, cm$ and $d_e\approx 10^{-44} \,e\,cm$ reflects, to our mind, the recent knowledge of $d_e$ within the SM. Recently in \cite{Chub16} another mechanism for $\mathcal{P},\mathcal{T}$-violating electron-nucleus interaction in atomic systems was suggested, namely exchange by Higgs boson with CP-violating electron vertex. This effect unlike the electron EDM contribution appears to be nonzero with 3-loop CP-violating vertex. If to use the value from \cite{Hoog90} for the 3-loop contribution to EDM, Higgs boson contribution is $d_e^{eqv,H} \approx 10^{-40} \,e\,cm$. On the other side if to obtain the 3-loop contribution from the four-loop result $10^{-44} \,e\,cm$, dividing it by the factor $\frc{\alpha_s}{4\pi}$, the Higgs boson contribution would be $d_e^{eqv,H} \approx 10^{-45} \,e\,cm$. In what follows we will adopt the ``benchmark'' value $d_e^{eqv} \approx 10^{-38} \, e\, cm$ as the maximum possible $\mathcal{P},\mathcal{T}$-odd effect in atomic systems in frames of SM.

We remind now the ordinary Faraday effect, having in mind a heavy atom with one valence electron in $ns_{1/2}$ state, located in an external magnetic field with the field strength $\mathcal{H}$. In more detail this theory was outlined in \cite{Nov77}, \cite{Rob80}. We assume that the light traveling through the atomic vapor in the external magnetic field has a resonance frequency
\begin{equation}
\label{1}
\omega = E_{np_{1/2}}-E_{ns_{1/2}}.
\end{equation}
In \Eq{1} $np_{1/2}$ is an excited atomic $p$-state; $E_{np_{1/2}}$, $E_{ns_{1/2}}$ are the energies of atomic levels. In an external magnetic field the levels are splitted by Zeeman effect in the following way \cite{Bethe57}:
\begin{equation}
\label{2}
E_{ns_{1/2}m}=E_{ns_{1/2}}^{(0)} +2\mu_0 m \mathcal{H},
\end{equation}
\begin{equation}
\label{3}
E_{np_{1/2}m}=E_{np_{1/2}}^{(0)} +\frac{2}{3}\mu_0 m \mathcal{H},
\end{equation}
where $E_{ns_{1/2}}^{(0)}$, $E_{np_{1/2}}^{(0)}$ are the energies in the absence of magnetic field and
\begin{equation}
\label{4}
\mu_0 =\frac{e\hbar}{2mc}
\end{equation}
is the Bohr magneton; $e,m$ are the charge (absolute value) and the mass of an electron, $c$ is the speed of the light and $\hbar$ is the Planck constant. By $\mathcal{H}$ is denoted the absolute value of the magnetic field strength and the quantum number $m$ is: $m=\pm \frac{1}{2}$.

In case of the observation along the direction of magnetic field the absorption spectral line $2s_{1/2} \rightarrow 2p_{1/2}$ is splitted in two lines with frequencies
\begin{equation}
\label{5}
\omega^+= E_{2p_{1/2,+1/2}}-E_{2s_{1/2,-1/2}},
\end{equation}
\begin{equation}
\label{6}
\omega^-= E_{2p_{1/2,-1/2}}-E_{2s_{1/2,+1/2}}.
\end{equation}
In the transitions with frequencies $\omega^{\pm}$ the light with right (left) circular polarization is absorbed. Therefore the absorption probabilities become different for the right (left) circularly polarized light. This leads to the Faraday effect, i.e. rotation of polarization plane for the linearly polarized light propagating through the atomic vapor. The difference $\Delta \omega =\omega^+ - \omega^-$ is proportional to the applied magnetic field
\begin{equation}
\label{7}
\Delta \omega=\frac{2}{3}\mu_0 \mathcal{H}.
\end{equation}
The absorption probability for the transition $ns\rightarrow np$ in the absence of magnetic field in the nonrelativistic limit is given by the standard formula
\begin{equation}
\label{8}
W=\frac{4}{3} \omega^3 e^2 | \langle ns | \boldsymbol{r}| np \rangle |^2
\end{equation}
where $\boldsymbol{r}$ is the radius-vector for the electron in an atom, $\omega$ is the transition frequency. In the weak external magnetic field (all the laboratory fields can be considered as weak in this respect) inserting the frequencies \Br{5}, \Br{6} in \Eq{8} we arrive at the difference between the probabilities of absorption for the right and left photons
\begin{equation}
\label{9}
\Delta W= W^+-W^-=4 \Delta \omega \omega_0^2 e^2 | \langle ns | \boldsymbol{r}| np \rangle |^2 = \frac{8}{3}\mu_0 \mathcal{H}\omega_0^2 e^2 | \langle ns | \boldsymbol{r}| np \rangle |^2
\end{equation}
where $\omega_0=E_{np_{1/2}}^{(0)}-E_{ns_{1/2}}^{(0)}$. Thus the Faraday effect is linear in external magnetic field.

Now we turn to the linear Stark shift of atomic levels caused by CP-violating phase in SM for an atom located in an external electric field $\bm{\mathcal{E}}$. In principle, the situation fully reminds the case with Zeeman effect in an external magnetic field: the absorption line is splitted again in two lines which correspond to the absorption of the right and left circularly polarized photons. The difference $\Delta W$ which determines the angle of polarization plane rotation is
\begin{equation}
\label{10}
\Delta W^{\mathcal{P},\mathcal{T}}=4 \Delta \omega^{\mathcal{P},\mathcal{T}} \widetilde{\omega}_0^2 e^2 | \langle \widetilde{ns} | \boldsymbol{r}| \widetilde{np} \rangle |^2
\end{equation}
where
\begin{equation}
\label{11}
 \Delta \omega^{\mathcal{P},\mathcal{T}}= \langle \widetilde{np} | S| \widetilde{np} \rangle - \langle \widetilde{ns} | S| \widetilde{ns} \rangle,
\end{equation}
$\langle ns(p) | S| ns(p) \rangle$ are linear Stark shift matrix elements for $ns$, $np$ levels. By $\widetilde{\omega}_0$ we denote the value of transition frequency modified by the external electric field with exclusion of the linear Stark shift corrections. The same concerns the notations $\widetilde{ns} \rangle$, $\widetilde{np} \rangle$ for the atomic wave functions. In the maximum available laboratory electric field $\mathcal{E} \sim 10^5 \;V/cm$ these modifications may be quite essential but not drastic and will not change our final estimates.

Expressions for the linear Stark matrix elements are essentially different in cases of the electron EDM effect and $\mathcal{P},\mathcal{T}$-odd electron-nucleus interaction in atomic systems. We begin with the electron EDM induced linear Stark effect (LSE). In this case Schiff theorem \cite{Schiff63} plays an important role. According to this theorem in the closed systems where the particles are bound by electrostatic forces, the total electric field acting on each particle including electrons is equal to zero. Hence, electron EDM effect should be absent. However this statement is violated in presence of non-electrostatic forces. In case of electrons Schiff theorem is violated by magnetic forces which are especially strong in heavy atoms, ions and molecules due to the strong relativistic effects. This leads even to the essential enhancement of the electron EDM effects in such systems \cite{San65}, \cite{Flam76}.

We present the electron EDM operator in the relativistic case as
\begin{equation}
\label{12}
\boldsymbol{\hat{d_e}}=d_e \boldsymbol{\Sigma}
\end{equation}
where $d_e$ is the absolute value for the electron EDM and $\boldsymbol{\Sigma}$ is the Dirac matrix. Then the expression for the linear Stark shift matrix element in an external electric field $\bm{\mathcal{E}}$ looks like \cite{Flam76}:
\begin{eqnarray}
 \label{13}
 \langle nl | S| nl \rangle_d &=& -d_e \langle nl | \left( \gamma_0-1 \right) \bm{\mathcal{E}} \boldsymbol{\Sigma} | nl \rangle
 \nonumber
\\
 &+& d_e e \bm{\mathcal{E}} \bigg[ \sum_{n'l'} \frac{\langle nl| \boldsymbol{r} |n'l' \rangle \langle n'l' | \left(\gamma_0-1 \right) \boldsymbol{\Sigma} \bm{\mathcal{E}_c} | nl \rangle }{E_{n'l'}^{(0)} - E_{nl}^{(0)}}
 \nonumber
 \\
  &+& \sum_{n'l'} \frac{\langle nl| \left(\gamma_0-1 \right) \boldsymbol{\Sigma} \bm{\mathcal{E}_c} |  n'l'\rangle \langle n'l' | \boldsymbol{r}|nl\rangle }{E_{n'l'}^{(0)} - E_{nl}^{(0)}} \bigg].
 \end{eqnarray}
In \Eq{13} $\bm{\mathcal{E}_c}$ is the Coulomb field of the nucleus
\begin{equation}
\label{14}
\bm{\mathcal{E}_c}= \frac{Ze\boldsymbol{r}}{r^3},
\end{equation}
$Ze$ is the charge of the nucleus. The presence of a factor $\left(\gamma_0-1 \right)$ in matrix elements in \Eq{14} retains in these matrix elements only the lower components of Dirac wave functions, i.e. pure relativistic contribution. This is a consequence of the Schiff theorem. In a similar way the contribution of the $\mathcal{P},\mathcal{T}$-odd electron-nucleus interaction can be presented \cite{Bon15}. The operator of this interaction according to \cite{Gor79} is
\begin{equation}
\label{15}
V_{\mathcal{P},\mathcal{T}}=iQ_{\mathcal{P},\mathcal{T}} g_{\mathcal{P},\mathcal{T}} \gamma_0 \gamma_5 \delta \left( \boldsymbol{r} \right)
\end{equation}
where $g_{\mathcal{P},\mathcal{T}}$ is the interaction constant, $Q_{\mathcal{P},\mathcal{T}}$ is ``$\mathcal{P},\mathcal{T}$-odd charge of the nucleus'', $\gamma_0, \gamma_5$ are the Dirac matrices. The value of $Q_{\mathcal{P},\mathcal{T}}$ depends on the particular model. In both models of this $\mathcal{P},\mathcal{T}$-odd interaction: ``2-photon benchmark'' model \cite{Pos14} and ``Higgs exchange'' model \cite{Chub16}, $Q_{\mathcal{P},\mathcal{T}}=A$ where $A$ is atomic number. Following \cite{Bon15} we present the linear Stark shift due to $\mathcal{P},\mathcal{T}$-odd electron-nucleus interaction in an external electric field as
\begin{eqnarray}
 \label{16}
 \langle nl | S| nl \rangle_{\mathcal{P},\mathcal{T}} &=& e \bm{\mathcal{E}} \bigg[ \sum_{n'l'} \frac{\langle nl| \boldsymbol{r} |n'l' \rangle \langle n'l' | V_{\mathcal{P},\mathcal{T}} | nl \rangle }{E_{n'l'}^{(0)} - E_{nl}^{(0)}}
 \nonumber
 \\
  &+& \sum_{n'l'} \frac{\langle nl| V_{\mathcal{P},\mathcal{T}} |  n'l'\rangle \langle n'l' | \boldsymbol{r}|nl\rangle }{E_{n'l'}^{(0)} - E_{nl}^{(0)}} \bigg].
 \end{eqnarray}
It is convenient to introduce the ratio
\begin{equation}
\label{17}
\eta_{nl} = \frac{\langle nl | S| nl \rangle}{\langle nl | - \boldsymbol{\mu_e}\boldsymbol{\mathcal{H}} | nl \rangle} =  \frac{\langle nl |- \boldsymbol{d_e}\boldsymbol{\mathcal{E}} | nl \rangle}{\langle nl | - \boldsymbol{\mu_e}\boldsymbol{\mathcal{H}} | nl \rangle}
\end{equation}
to compare the $\mathcal{P},\mathcal{T}$-odd and ordinary Faraday effects in comparable $\boldsymbol{\mathcal{E}}$ and $\boldsymbol{\mathcal{H}}$ fields. Instead of $d_e$ in \Eq{17} the value $d_e^{eqv}$ also can be inserted as it was explained above. Evaluation of $d_e^{eqv}$ for the given $ V_{\mathcal{P},\mathcal{T}}$ see, for example in \cite{Chub16}. Employing the Gauss system of electromagnetic units where the electric $\boldsymbol{d_e}$ and magnetic $\boldsymbol{\mu_e}$ dipole moments (as well as electric and magnetic field strengths) have the same dimensionality we can write
\begin{equation}
\label{18}
d_e = | \boldsymbol{d_e} | =\mu_0 \xi_d = 1.68 \cdot  10^{-11} \xi_d \, e\, cm
\end{equation}
where $\xi_d$ is a dimensionless constant. From the most stringent recent experimental bound \cite{ACME13} it follows that
\begin{equation}
\label{19}
\xi_d < 10^{-18}.
\end{equation}
The largest predicted theoretical value within SM gives
\begin{equation}
\label{20}
\xi_d \approx 10^{-27}.
\end{equation}
Polarization plane rotation angle in the ordinary Faraday effect can be estimated as
\begin{equation}
\label{21}
\varphi (rad) = \mathcal{K} l (cm) \mathcal{H} (Oe).
\end{equation}
Coefficient $\mathcal{K}$ in \Eq{21} is different for different media. Its average value is about $\mathcal{K} \approx 10^{-3}$. For the $\mathcal{P},\mathcal{T}$-odd Faraday effect employing the ratio
\begin{equation}
\label{21a}
\eta_{nl} \approx \frac{\mathcal{K_{\mathcal{P},\mathcal{T}}}}{\mathcal{K}} \xi_d \frac{\mathcal{E} (V/cm)}{\mathcal{H} (Oe)}
\end{equation}
we can write
\begin{equation}
\label{22}
\varphi_{\mathcal{P},\mathcal{T}} (rad) = \mathcal{K}_{\mathcal{P},\mathcal{T}} \xi_d l (cm) \mathcal{E} (V/cm).
\end{equation}
The coefficient $\mathcal{K}_{\mathcal{P},\mathcal{T}}$ is defined by Eqs \Br{13} or \Br{16}. The matrix elements in Eqs \Br{13}, \Br{16} should be proportional to $Z^3$ as for the $\mathcal{P}$-odd effects in heavy atoms, where $Z$ is the charge of the nucleus \cite{Khrip91}. Assuming that the coefficient $\mathcal{K}_{\mathcal{P},\mathcal{T}}$ is also proportional to the coefficient $ \mathcal{K}$ in \Eq{21} we obtain the maximum value for $\mathcal{K}_{\mathcal{P},\mathcal{T}}$ for heavy atoms ($Z \sim 10^2$):
\begin{equation}
\label{23}
\mathcal{K}_{\mathcal{P},\mathcal{T}} \approx 10^{-3} \cdot Z^3 \approx 10^3.
\end{equation}
The value of $l$ in \Eq{22} is limited by the absorption of the light in the medium. We adopt the maximum value of $l$ as $\sim 70000  \, km=10^{10} \, cm$ \cite{Baev99}. Then, applying maximum possible laboratory electric field $\mathcal{E}\approx 10^5 \, V/cm$ we obtain the maximum rotation angle within SM
\begin{equation}
\label{24}
\varphi_{\mathcal{P},\mathcal{T}}  \approx 10^{-9} \, rad.
\end{equation}
This value is within the recent experimental possibilities. Comparing the $\mathcal{P},\mathcal{T}$-odd Faraday effect measurement with the modern experiments on the $\mathcal{P},\mathcal{T}$-odd effects in atoms and molecules where electron spin precession in an external electric field is measured \cite{Reg02}-\cite{ACME13}, we admit that the Faraday experiment is less sensitive to the most dangerous false effect originating from the motional magnetic field. In the Faraday experiment a systematic motional magnetic field effect is absent since the propagating photons are neutral. The motional false $\mathcal{P},\mathcal{T}$-odd effect may arise only as statistical one due to the chaotic motion of the neutral atoms (molecules) in the medium in external electric field. Systematic effects can arise only due to parasitic external magnetic fields, e.g.  field of the Earth. In this sense the situation is the same as for the experiments on the electron spin precession in external electric field. It should be mentioned also how it is possible to distinguish the $\mathcal{P},\mathcal{T}$-odd Faraday effect from the much larger $\mathcal{P}$-odd optical rotation effect. The answer is that the latter one like the ordinary optical activity effect in optically active media cancels when the light travels onward and backward in the cavity. Contrary to this the $\mathcal{P},\mathcal{T}$-odd Faraday effect like the ordinary Faraday effect grows up during the onward and backward travels.

\textbf{Acknowledgements}

This work was supported by RFBR grant 14-02-00188 and by the St. Petersburg State University grant 11.38.227.2014. The work of D.C. was also supported by the non-profit foundation ``Dynasty'' (Moscow).

\end{document}